\documentclass[twocolumn,showpacs,preprintnumbers,amsmath,amssymb]{revtex4}
\usepackage{tabularx,graphicx}\begin{document}
\newcommand{\beq}{\begin{equation}}
\newcommand{\eeq}{\end{equation}}
\newcommand{\beqn}{\begin{eqnarray}}
\newcommand{\eeqn}{\end{eqnarray}}
\newcommand{\bmath}{\begin{subequations}}
\newcommand{\emath}{\end{subequations}}

\title{ The Lorentz force and superconductivity}
\author{J. E. Hirsch }
\address{Department of Physics, University of California, San Diego\\
La Jolla, CA 92093-0319}

\date{\today} 

\begin{abstract}
To change the velocity of an electron requires that a Lorentz force acts on it, through an
electric or a magnetic field. We point out that within the conventional understanding
of superconductivity electrons appear to change their velocity in the absence of Lorentz forces.
This indicates a fundamental problem with the conventional theory of 
superconductivity. A hypothesis is proposed to resolve this difficulty.
This hypothesis is consistent with the theory of hole superconductivity. \end{abstract}
\pacs{}
\maketitle 

In the microscopic realm, electrons do not change their state of motion in the
absence of an electromagnetic force (we omit consideration of gravitational
forces throughout this paper). For example, in the Stark effect the electron changes
its wavefunction when an electric field is applied because of the electric Lorentz force acting on the electron. 
In  paramagnetic atoms, orientation of atomic magnetic moments under application of a
magnetic field can be understood as arising from the magnetic Lorentz force
on orbital or intrinsic (due to spin) electric currents.
 In a diamagnetic atom, the wavefunction of the electron does
not change upon application of a magnetic field (to lowest order) but its velocity does. From the relation between
velocity and canonical momentum $\vec{p}$ in the presence of a magnetic vector potential $\vec{A}$,
\beq
\vec{v}= \frac{\vec{p}}{m_e}-\frac{e}{m_ec}\vec{A}
\eeq
one finds that the change in velocity when the magnetic field is increased from zero to its finite value is
\beq
\Delta\vec{ v}=-\frac{e}{m_ec}\vec{A}
\eeq
since the wavefunction and consequently the canonical momentum do not change to first order\cite{slater}.
Here, $m_e$ is the free electron mass. This change of velocity can be understood as arising from the
Lorentz electric force acting on the electron, with the electric field generated by the
changing magnetic field through Faraday's law\cite{slater}. I argue that we know of no example
in the microscopic quantum-mechanical world where electrons would change their
velocity in the $absence$ of an applied Lorentz force
\beq
\vec{F}=e(\vec{E}+\frac{\vec{v}}{c}\times\vec{B})
\eeq
which requires either an electric field, or a magnetic field together with a non-zero velocity.
 
In the macroscopic world, new phenomena may occur when many degrees of freedom are 
at play\cite{anderson}. For example, a ferromagnet will 'spontaneously' develop a magnetic moment
when cooled below its critical temperature. Still, even in that case no net magnetic moment will
be observed in the absence of an applied magnetic field because of domain formation.
If the metal is cooled in the presence of a magnetic field the magnetic moments of the domains
will orient in the direction of the external field because of the torque that a magnetic moment
experiences in the presence of a magnetic field, which we can also
attribute to  the Lorentz force Eq. (3).

In contrast, in superconductors new phenomena are observed that appear to be more
'mysterious' than those seen in both the microscopic world as well as in other
macroscopic phase transitions: electrons change their state of motion in certain
specific ways that appear to be independent of forces acting on the electrons. I argue that
these situations present a puzzle within the conventional theory of superconductivity.
An explanation of these phenomena is proposed that requires a revision of the
conventional understanding of supeconductivity.

\subsection{The Meissner effect}
The expulsion of magnetic flux from the interior of a type I superconductor when a
magnetic field is turned on can be understood, just as diamagnetism of atomic electrons,
as arising from the current generated through the Lorentz electric force generated by
the changing magnetic field acting on the superfluid electrons\cite{becker}.
However when a simply connected  type I superconductor is cooled below $T_c$
in the presence of a constant magnetic field, 
the same final state needs to be achieved, in the absence of electric forces generated
by changing magnetic fields. Furthermore since the superfluid is supposed to
be at rest, no magnetic Lorentz force can act. How is it possible then that a state with
finite screening currents is reached?

London postulates the London equation\cite{london}
\beq
\vec{v}_s=-\frac{e}{m_ec}\vec{A}
\eeq
for the superfluid velocity that should exist in the presence of the vector potential $\vec{A}$,
$regardless$ of how that state was reached. No justification for Eq. (4) exists within
the standard theory of electromagnetism. As a plausibility argument it is argued that
the wavefunction of the superconductor is 'rigid' and that Eq. (1) with $\vec{p}=0$ applies for a 
simply connected superconductor independent of history, so that Eq. (4) 
is the $only$ possible state due to this rigidity.
However this begs the question: rigidity or not, in the atom the diamagnetic
current is generated through
a real force generated by a real electric field acting on the atomic electron. Why can't
the final state of the superconductor be understood in the same way?
How do the electrons in the superconductor 'know' to start moving when the metal is
cooled below $T_c$ in the presence of a magnetic field?

\subsection{The rotating superconductor}

In a simply connected superconductor rotating with angular velocity $\vec{\omega}$,
a magnetic field exists throughout its interior given by\cite{becker,london}
\beq
\vec{B}=-\frac{2m_e c}{e}\vec{\omega}
\eeq
(conventionally called 'London field')\cite{londonmy}.
This has been verified experimentally for both conventional\cite{pb,nb,sn}
 and high $T_c$\cite{bapb}
superconductors. Theoretically it was predicted first based upon the theory
of perfect conductors\cite{becker}, for the case where the metal in the superconducting state
is put into rotation. In that framework it can be understood as arising because the electrons near
the surface 'lag behind' when the body is put in rotation, and a surface current
is generated.   When the ions start moving the resulting electric current
due to ionic motion generates a changing magnetic field which in turn generates
an electric field that makes the electrons follow suit.

The existence of the field Eq. (5) also follows from London's equation\cite{london}, and hence is
predicted to exist also when a rotating normal metal is cooled below its
superconducting transition temperature, and indeed is so found experimentally\cite{sn}.
However we face then a similar problem as for the Meissner effect discussed above. If the
electrons are rotating with the lattice in the normal state, what makes the electrons
near the surface
'lag behind' when the metal becomes superconducting to generate the interior
magnetic field Eq. (5)? No magnetic field nor electric
field should initially exist, so no Lorentz force acts on the electron.

Furthermore there is another mysterious consequence of Eq. (5). In the interior of the
superconductor the electrons are rotating at the same angular velocity as the lattice.
Assuming no force is exerted by the ionic lattice on the superfluid, the centripetal
force for the electron to rotate needs to be provided by the magnetic field. However
the magnetic field required for a charge $e$ and mass $m_e$ to rotate   with angular velocity $\omega$ is
$half$ the value of the magnetic field Eq. (5)! In other words, an electron rotating
in a magnetic field rotates at the cyclotron frequency $\omega=eB/m_ec$ rather than
the Larmor frequency $\omega=eB/2m_ec$. Consequently, for mechanical 
equilibrium, an electric
field in the interior of the superconductor needs to exist:
\beq
\vec{E}=\frac{\vec{B}\times{(\vec{\omega}}\times\vec{r})}{2c}
\eeq
\begin{figure}
  \includegraphics[height=.25\textheight]{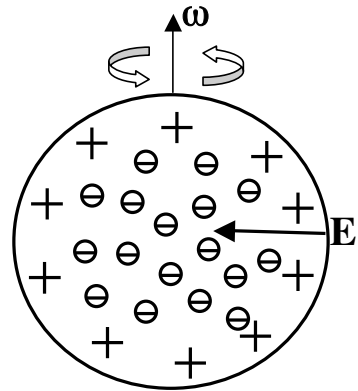}
  \caption{Charge configuration in a rotating superconductor implied by the conventional theory (qualitative). 
For mechanical equilibrium, an electric field pointing in needs to exist, generated by negative electrons moving
in slightly giving a non-uniform charge distribution.}
\end{figure}
 This electric field points towards the $interior$ of the superconductor. Hence it requires the negative
superfluid to move slightly $in$ towards the interior of the superconductor to generate this field,
as shown schematically in Figure 1. However, one would expect exactly the opposite:
if in a rotating metal the electrons become 'free' as the metal enters the superconducting state, the
centrifugal force would push the electrons $out$ rather than in. However if such was the case the
resulting electric field would point $out$, which would be incompatible with Eq. (5) and mechanical
equilibrium.

We should point out that previous discussions of the electric field inside rotating superconductors 
within the conventional framework erroneously concluded that an electric field pointing 
$out$ exists\cite{gaw,rys},
compatible with the expectation\cite{becker}  that electrons should
move $out$ due to the centrifugal force. This is because in analyzing
the situation in the rotating frame, the contribution to the electric field arising from Lorentz-transforming
the magnetic field Eq. (5) was omitted.

\begin{figure}
  \includegraphics[height=.25\textheight]{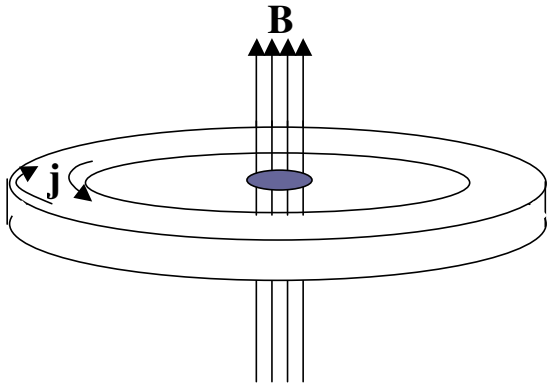}
  \caption{Superconducting ring  threading magnetic flux. The magnetic field lines are confined to a small central
region far away from the inner surface of the ring. No magnetic field exists anywhere in the ring. A current {\bf j} will exists near
the ring surfaces if the applied flux is not an integer multiple of the flux quantum.}
\end{figure}

\subsection{The quantized flux}
Consider a metal ring with magnetic flux through its center in a well-localized region that does not overlap
the ring, as shown in Figure 2. The flux quantization condition\cite{london}
\beq
\oint \vec{p} \cdot \vec{dl}=nh
\eeq
($n$=integer, $h$=Planck's constant)
requires that if the ring is in the superconducting state the magnetic flux
enclosed is an integer multiple of the flux quantum\cite{onsager}:
\bmath
\beq
\Phi=\int  \vec{B}\cdot \vec{dS}=n\Phi_0
\eeq
\beq
\Phi_0=\frac{hc}{2e}
\eeq
\emath
If the applied magnetic field does not satisfy the condition Eq. (8), surface currents develop in the ring so that
Eq. (8) is satisfied. If the external magnetic field is changed from a value that satisfies Eq. (8) to one that does not,
while the ring is in the superconducting state, the development of these ring currents can be understood:
as the magnetic flux is changed, magnetic field lines will move across the superconducting ring and exert a
 Lorentz force on the superfluid electrons that will drive the ring surface currents necessary to 
satisfy Eq. (8).

However, if the ring is cooled from the normal to the superconducting state while enclosing a flux that does not
satisfy Eq. (8), how do the ring currents develop? In that case no magnetic field ever exists in the ring itself, as well as no electric field, so the Lorentz force is zero. How do the
electrons know to start moving?

We argue that the three examples discussed above represent unsolved puzzles in the
conventional understanding of superconductivity. In the following we propose a hypothesis that
explains these puzzles.  

A hint to explain these puzzles arises from consideration of the electron in a diamagnetic 
atom. In the atom, it is the $change$ in the electron velocity that obeys the
London-like  equation (2). This can be simply understood classically. Assume the
electron is rotating in an orbit of radius $r$. The centripetal force is provided by the
ionic electric field $E_{ion}$:
\beq
\frac{m_ev^2}{r}=eE_{ion}
\eeq
On applying a magnetic field, the $change$ in the centripetal force is provided
by the magnetic Lorentz force. In absolute value,
\beq
\frac{2m_e v \Delta v}{r}=\frac{ev}{c}B
\eeq
leading to
\beq
\Delta v=\frac{e}{m_ec}\frac{Br}{2}
\eeq
which is equivalent to Eq. (2) for $\vec{A}=\vec{B}\times \vec{r}/2$. The left-hand side of 
Eq. (10) follows from a variation of Eq. (9) only if $\Delta v << v$, which is consistent 
with the fact that quantum-mechanically Eq. (2) only holds to first order in the
magnetic field.

From an identical consideration it is clear that we will have mechanical equilibrium for 
the superfluid electrons in the rotating superconductor with the correct factor of 
$2$ in the London field Eq. (5) {\it if the superfluid electron is already rotating at a high
angular velocity before the body is set into rotation}, so that Eq. (10) applies just as for
the electron in the atom (from Eq. (10), Eq. (5) follows for 
rigid rotation with $\Delta v=\omega r$).  If this is so, an electric field
has to exist in the interior of the superconductor which provides the centripetal
force to sustain the superfluid electron rotation. This implies that a non-zero $positive$ charge
density exists in the interior of the superconductor, which in turn leads us to 
conclude that negative charge is expelled from the interior of the metal towards the
surface when the metal enters the superconducting state.
 
\begin{figure} \includegraphics[height=.22\textheight]{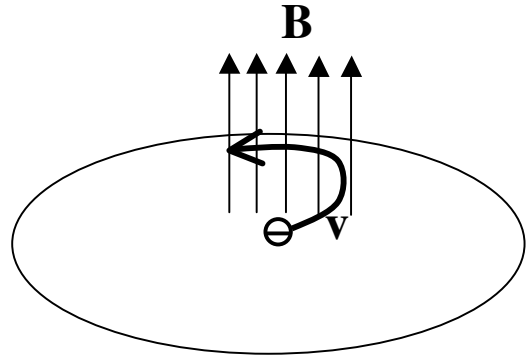}

  \caption{Qualitative explanation of the Meissner effect. As electrons are expelled from the 
interior of the superconductor, the radially outgoing velocity in the presence of the B-field give rise to 
a tangential Lorentz force that drives the screening current.}
\end{figure}

Remarkably, this hypothesis then   provides us with an explanation of the Meissner
effect. When the system goes superconducting electrons in the interior are expelled
towards the surface. In the presence of a magnetic field, the
Lorentz force on the radially outgoing electron will give rise to a tangential force in the direction
needed to generate the surface currents that will screen the magnetic field, as shown 
schematically in Figure 3.

Furthermore this assumption provides a natural explanation for why the superfluid electrons
 near the surface 'lag behind' when a rotating metal becomes superconducting: 
 electrons flow out, and if the body is rotating there is a
Coriolis force on the outward flowing electrons that makes them lag behind when
they approach the surface. Simply put, the electron from the interior has a smaller tangential velocity
so that when it flows out it will lag the faster motion occuring at larger r.

\begin{figure}
 \includegraphics[height=.25\textheight]{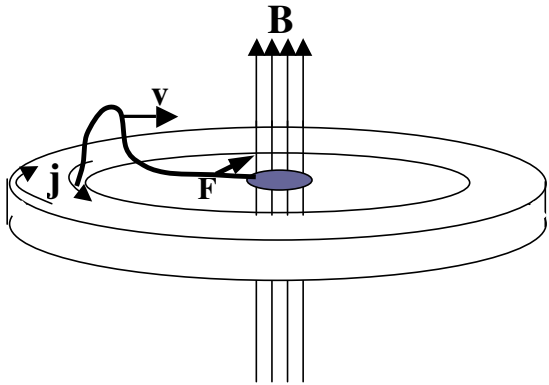}
\caption{Explanation of the flux quantization puzzle. When the superconductor expels electrons from its interior,
the electronic wave function 'leaks out' of the body of the superconductor.   The tail of the electronic wave function has
to extend into the region where the magnetic field is nonzero to feel the Lorentz force and start moving.}
\label{fig6}
\end{figure}

Finally, how do the electrons in the superconducting ring threaded by magnetic flux 
get set into motion when
neither a magnetic field nor an electric field nor a  preexistent superfluid velocity exist in the ring? When the 
superconductor expels electrons from its interior we need to assume that the wavefunction
of these electrons near the surface 'leaks out' and reaches the region where the
magnetic field is non-zero, as shown in Figure 4. This occurs through a radially inward velocity, which when the
wavefunction reaches the region of non-zero magnetic field
 gives rise to a Lorentz force in the tangential direction that can set the surface 
current in the superconducting ring into motion. Note that this implies that the current 
generated will always be such that the magnetic field generated by the ring currents opposes the pre-existing flux, hence will 'round down' the non-integral flux quantum number generated
by the applied field.

We point out  that the theory of hole superconductivity\cite{hole} predicts that
electrons are expelled from the interior of superconductors when the transition to
superconductivity occurs\cite{charge}, that a radially outward electric field exists in the
interior of superconductors\cite{charge,atom}, and that the wavefunctions of electrons will leak out
from the body of the superconductor\cite{atom}, as required by the explanations discussed above.
Of course other explanations may also be possible.

The reader may note that the proposed solution to these puzzles raises another puzzle.
If superfluid electrons are rotating in the absence of body rotation and magnetic fields, why is no
current observed? The reason is that when electrons are expelled from the interior of the
superconductor, interaction of the electron spin of the radially outgoing electron with the ionic
lattice will deflect electrons of opposite spin tangentially in opposite directions\cite{atom,ferro}.
As a consequence, macroscopic spin currents are predicted to exist in superconductors if this
scenario is correct. This phenomenon and some experimental consequences are also discussed in 
ref. \cite{atom}.

 \end{document}